\documentclass[twocolumn, notitlepage,aps,showpacs,floats,amssymb,amsmath,floatfix,groupedaddress,superscriptaddress,aps,pre,longbibliography]{revtex4-2}
\usepackage{amsfonts,amssymb,stmaryrd,latexsym,amsmath,braket}
\usepackage{graphicx} 
\usepackage{booktabs}
\usepackage{comment}
\usepackage{newtxtext, newtxmath} 
\usepackage{slashed}
\usepackage{bm}
\usepackage{appendix}
\usepackage{enumitem}
\usepackage{multirow}
\usepackage{array}

\usepackage{placeins}
\usepackage[dvipsnames]{xcolor}

\usepackage[colorlinks=true,linktocpage=true,citecolor=blue,urlcolor=blue,linkcolor=blue]{hyperref}
\begin{document}

\title{Effects of Charge Regulation on the Conformations and Dynamics of a Polyelectrolyte during  Nanoparticle Adsorption}

\author{Kawaljeet Kaur}
\affiliation{Department of Physics, Indian Institute of Technology Jodhpur, Jodhpur, Rajasthan 342030, India}

\author{Rashmi Kandari}
\affiliation{Department of Physics, Indian Institute of Technology Jodhpur, Jodhpur, Rajasthan 342030, India}

\author{Sunita Kumari}
\email{sunita@iitj.ac.in}
\affiliation{Department of Physics, Indian Institute of Technology Jodhpur, Jodhpur, Rajasthan 342030, India}

\author{Subhajit Paul}\email{spaul@physics.du.ac.in, subhajitphys@gmail.com}
\affiliation{Department of Physics and Astrophysics, University of Delhi, Delhi 110007, India}

\date{\today}

\begin{abstract}
Customizing the surface characteristics and stimuli-responsive behavior of polyelectrolyte (PE)  during interaction with nanoparticles (NP) have potential applications ranging from biological interactions to advanced diagnostic technologies. Using numerical  simulations, we investigate how charge regulation (CR) shapes the conformation, dynamics  and adsorption of PE on a oppositely charged NP and compared the results with constant charge (CC) conditions. At low solvent ion concentrations, CR barely affects the adsorption fraction, while significantly enhancing the dynamical response of the PE. However, at high ion concentrations, while complexation is suppressed in the CC case, the CR environment helps to overcome the energy barriers related to Debye screening and thermal fluctuations, thereby promoting the adsorption. By maintaining the strong  PE-NP electrostatic interaction, PE exhibits a more directed ballistic-like dynamics towards NP during adsorption, in comparison to the diffusive and sub-diffusive behavior observed in the CC case. Our results emphasize that in high ion concentrations, with a strong Debye screening effect, a small fraction of charge-fluctuating sites suffices for adsorption to happen. A tunable CR mechanism is thus an efficient and robust strategy to control the biomolecular interaction and stability in complex environments. \\


\end{abstract}

\maketitle

\section{Introduction}
\label{sec:intro}
The adsorption of charged polymers or polyelectrolytes (PEs) onto charged surfaces remains a cornerstone in many aspects of our lives \cite{Muthu2023,Dobrynin2001,Netz2003}. 
Such processes can mitigate colloidal aggregation  \cite{Netz2003,Rubinstein2012,Yuan2026}, nanoparticle templating  \cite{ulrich2005,Dotzauer2006,Roshan2025}, engineering of multilayer films for sensors and catalysis, etc. Understanding of many biological processes also, interaction between biopolymers (for example, DNA) and protein are important  \cite{Cooper2005,Samanta2020,Stornes2021}. The orientational specificity of proteins adsorbed on charged surfaces is governed by their heterogeneous charge distributions, which create oppositely charged patches that drive directional electrostatic interactions \cite{Anhesini2026}. In the biomedical and nanotechnology arenas, stimuli-responsive PE-surface interactions are a key ingredient for targeted delivery, controlled release, and encapsulation\cite{Delcea2011,Kataoka2001}. Considerable efforts have been made to understand such PE adsorption on surfaces\cite{Cherstvy2012repulsion, Africo2024}, although these efforts primarily rely on solving the Poisson equation under the assumption of constant charge (CC) for both PE and the adsorbing surfaces\cite{kesson1989, podgornik_colloidal_1995, Miklavic1990, Gurovitch1999, Vagharchakian2004,Szilagyi2014,Changizrezaei2026}. However, in response to local environmental signals, both PE and surface dynamically adjust their ionizable surface groups \cite{Gummel2007, Samanta2020, Muthu2023, Kumari2024, Yuan2024, Beyer2025, Beyer2026, Radhakrishnan2025, Kandari2026, Yadav2026, Yuan2026}. This  charge control mechanism is termed charge regulation (CR) and often leads to intriguing phenomena \cite{Blanco2019,Stornes2021, Samanta2020, Muthu2023, Kumari2024, Yuan2024, Beyer2025, Kandari2026, Yadav2026, Yuan2026, Beyer2026,Muthukumar2010, Samanta2020,Radhakrishnan2025}. For example, by reducing electrostatic repulsion between similarly charged PE and the surface, CR can enhance the  PE adsorption compared to CC approximations \cite{Yuan2024,Yadav2026}. Moreover, CR includes many-body phenomena dictated by local charge fluctuations (charge capacitance), which are especially significant for self-assembly with heterogeneous or patchy surfaces\cite{Yigit2015,Radhakrishnan2025},  pH dependent adsorption and force modulation between cationic PE and silica surfaces \cite{huang}. 
Fully charged PEs can be fully or partially wrapped around the NP depending on the surface chemistry of the NP and the concentration of the solution\cite{ulrich2005, ulrich2006}. Simialrly,   Stornes et al. elucidate that with large difference between the acid and base dissociation constants, CR dominates the PE–NP complexation process by significantly altering the adsorption behavior, stability, and the structural properties \cite{Stornes2021}. In another study, multi–polyelectrolyte–protein complexation has been investigated and CR has been shown to promote more compact complex structures \cite{Samanta2020}.



The interaction force between two similarly charged objects is primarily controlled by strong repulsion arising from the overlap of diffusive double layers formed by counterions covering the charged surfaces \cite{chapman1913, derjaguin1941, israelachvili_intermolecular_2011}.The classical DLVO theory, developed by Derjaguin, Landau, Verwey and Overbeek, describes the stability of colloidal systems through the competition between attractive van der Waals interactions and repulsive electrostatic double-layer interactions\cite{Verwey}. However, this does not consider the electrostatic ion-ion and ion-surface site correlations and also breaks down in case of charge regulation (CR) scenario.  The concept of CR was first introduced by Linderstrøm-Lang regarding ionization properties of proteins \cite{lang}. Forces arising due to such charge fluctuations was formally  calculated by Kirkwood-Shumaker \cite{wood52}. However, difficulties arise for complex charged macromolecules or systems exhibiting multivalency. Ninham and Parsegian provided the qualitative description of the CR effect, by incorporating variable surface charge densities using a nonlinear Poisson-Boltzmann (PB) theory for planar surfaces \cite{Ninham1971}. Following this, both the law of mass action \cite{1975regulation, 1976surface, PericetCamara2004, von1999} and surface free energy approaches \cite{long2012, 2014field, 2015charge, 2015surface, 1996kin} were used to describe CR by modeling surface binding sites. Using mean-field arguments, Muthukumar proposed  a theoretical framework for understanding polyelectrolyte (PE) adsorption onto a charged surface \cite{Muthukumar1987}. However, the complexity of the solution and discrete nature of the ions limits the applicability of the mean-field PB picture\cite{Muthukumar2010, Stornes2021}. These parameters are considered to play a crucial role in determining the CR condition\cite{ulrich2005, ulrich2006}. 
In a few recent works, the effect of CR is only considered for the PE, while the NP is modeled as comprising patches of positive and negative fixed-charge ions \cite{Samanta2020}. Recently, a new simulation method has been proposed for the  modeling of titration curves which works for both semi-grand canonical and canonical ensembles\cite{Levin2023, Colla2024}. Recent works demonstrated that CR increases adsorption on similarly charged surfaces by charge suppression, whereas in oppositely charged systems, cooperative charge amplification further stabilizes binding \cite{Yuan2024}. When both the PE and the spherical surface are charge regulated, the interaction becomes highly coupled due to feedback between the polymer conformation and ionization. Simulations of PE-NP systems indicate that mutual CR promotes adsorption relative to CC models and can mitigate traditional bridging-induced attractions, substituting them with weaker osmotic or fluctuation-mediated interactions \cite{Yadav2026, Stornes2021}.

Even though, the CR case is considered in simulations to mimic more realistic experimentally relevant situations and considered its effect mostly on the adsorption fraction and forces due to the charge fluctuations, the conformational properties or the dynamics of the PE are not thoroughly looked at.  In this work, considering the effects of CR on both NP and PE, we investigate how CR modifies the dynamics during interactions between them, highlighting the role of CR in different ion concentrations. We compare these results with the conventional CC paradigm. To include the CR effect during simulation, we consider a hybrid Monte Carlo/Molecular dynamics (MC/MD) approach \cite{Curk2022,Curk2021}. We present interesting results  related to the scaling behavior of mean-squared-displacement (MSD) of the center-of-mass of PE during its approach towards the NP surface. In comparison to the diffusive or sub-diffusive scaling of the MSD for the CC case, the dynamics becomes much faster mostly following a ballistic scaling in the CR case. Our model allows all components to dynamically alter their charge in response to environmental signals.

The paper is organized as follows. The details of the model and the simulation methodology is presented in Sec.~\ref{sec:model}. Next, in Sec.~\ref{sec:result}, we present the  results for both low- and high-concentration limits, followed by a summary and conclusion in Sec.~\ref{sec:con}.

\section{Model and Methods}
\label{sec:model}

\begin{figure}
    \centering
    \includegraphics[width=0.5\textwidth]{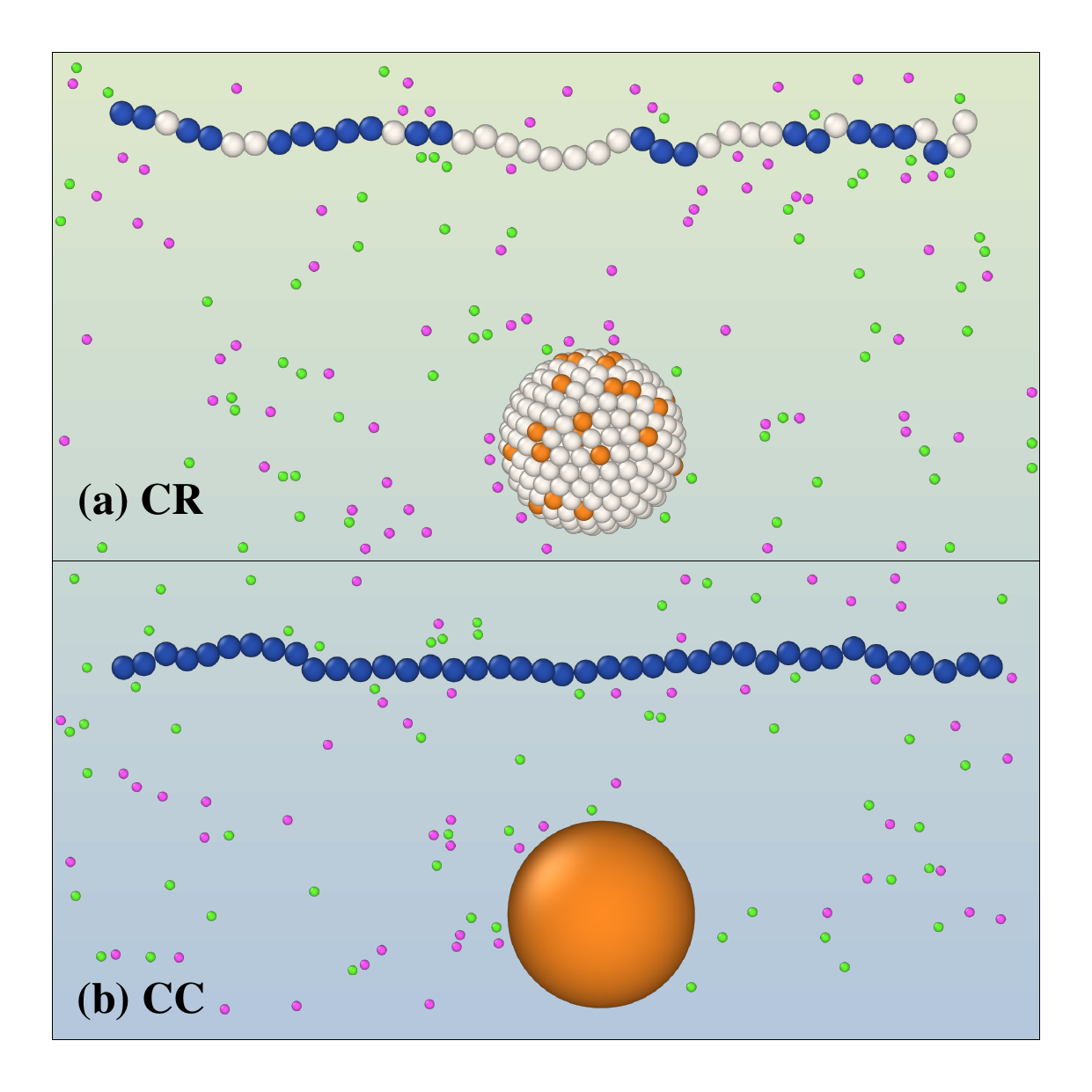}
    \caption{Schematics of the starting configurations of our model for charge regulated (CR) in (a) and constant charge (CC) case in (b). Nanoparticle (NP) of radius $R=5\ell_B$ is made up of $256$  surface groups. For the CR case, only $50\%$ fraction of the surface groups are basic (orange). The surface sites marked white correspond to neutral groups. For the polyelectrolyte (PE) also out  of $N$ monomers $50 \%$ are the charged (acidic/blue) ones.  Green and magenta particles represent positive and negatively charged ions in the solution, respectively. For CC case, the NP as well as each monomer of the PE are assigned fixed charge equal to the average charge obtained from the CR simulation.}
    \label{fig:Schematic}
\end{figure}
In Fig.~\ref{fig:Schematic} we show the schematic of our model, i.e., a NP of radius $R$ and an oppositely charged PE with $N$ monomers. The system is immersed in a monovalent electrolyte solution (e.g., salt, protons, hydroxyl ions, etc.) maintained at $pH = 7$. Although the bulk environment remains neutral, various surface interactions can dramatically alter the local chemical environment near the surface. In our simulation, $pIp$ and $pIm$ are logarithmic representations of the effective concentrations of combined cationic and anionic species, respectively. Both are defined in the  $-\log_{10}$  representation. For example, $10^{-{pIp}} = 10^{-{pH}} + 10^{-{pSp}}$, where $pSp$ denotes the concentration of added salt cations \cite{LAMMPS}.

The surface of the NP is covered with 256 ionizable  sites, characterized by a base dissociation constant $pK_b$. Among these surface groups, we assume that only $50\%$ are basic. The symmetric distribution of surface sites is generated according to the electron distribution in the Thomson problem\cite{tp06}.  The PE is modeled using a flexible bead-spring polymer chain of $N$ monomers with acidic groups with dissociation constant $pK_a$. Here, an  acidic site refers to the case which can acquire charge $q = 0$ or $-1]$. Similarly, a basic site can acquire $q=0$ or $+1$.
During simulation, these charged sites can reversibly alter their charge states via Monte Carlo (MC) moves, while particle positions are updated according to velocity-Verlet molecular dynamics (MD).
The acidic or basic ionizable sites follow the reactions,  $A\overset{{pk_a}}{\rightleftharpoons} A^- + H^+$ and $B\overset{pk_b}{\rightleftharpoons} B^+ + OH^-$, respectively \cite{Curk2021}. For simplicity, we assume that the monomers, ionizable sites, and mobile ions are identical in their sizes with diameter  $\sigma=\ell_B$,  where $\ell_B = e^2/(4\pi \varepsilon\varepsilon_0k_B T)$ is known as the Bjerrum length and $\varepsilon$ and $\varepsilon_0$ correspond to the dielectric constants.  This represents the distance at which the electrostatic interaction between two units of charges becomes equal with the thermal energy $k_B T$.  Such hybrid MC–MD simulations are performed  in LAMMPS\cite{LAMMPS}. After initial energy minimization and equilibration, production involves interleaved MD integration and MC trial moves that attempt protonation/deprotonation of titratable groups. Specifically, for each fixed interval of timesteps, an MC titration routine is invoked for both NP  and monomers. We have also performed CC simulations with the same system parameters as in the CR case. We first equilibrate a CR run and compute time-averaged charges on each titratable site over the last $5\times10^4$ MD steps. These fixed charges are then used in a CC run starting from same equilibrated structure. Thus, CR and CC simulations differ in whether ionization or charge fluctuations are allowed dynamically. In both CC and CR cases, the position of the  NP is kept fixed.

The input value of $pH$ determines the proton chemical potential of the external reservoir. The simulations are performed in a cubic box of length $L = 200 \ell_B$, with periodic boundary conditions in all directions. We used a fixed value of $\ell_B = 0.72$ nm for water at room temperature in all simulations. 

The nonbonded steric interactions between all particle pairs are given by the standard repulsive Lennard-Jones(LJ) potential as
\begin{equation}
    U_{\text{LJ}}(r_{ij})=
    \begin{cases}
      4\epsilon \left[ \left(\frac{\sigma}{r_{ij}-\Delta}\right) ^{12}  - \left(\frac{\sigma}{r_{ij}-\Delta}\right) ^{6} \right] , & r_{ij} \leq r_c~,\\
      0~, & r_{ij}>r_c~,
    \end{cases}
    \label{lj}
\end{equation}
where $r_{ij}$ is the distance between the particles with $\Delta= R-\sigma/2$ is the expanded distance and the cut-off distance $r_c=\Delta + 2^{1/6}\sigma$, $R$ is the radius of the NP and $\epsilon$ is the depth of the potential well. Bonded interactions between adjacent monomers in the PE are modeled with a harmonic potential
\begin{equation}
U_{\mathrm{bond}}(r) = \frac{1}{2}k (r - b_0)^2,
\end{equation}
with spring constant $k =400$ chosen to keep the bonds close to the equilibrium length $b_0 =1.5 \times 2^{1/6} \sigma$.




Long-range electrostatic interaction between charges $q_i$ and $q_j$ at positions $\vec{r}_i$ and $\vec{r}_j$ is described by the Coulomb potential
\begin{equation}
U_C(r_{ij})=
\begin{cases}
 \frac{1}{4\pi \varepsilon_0 \varepsilon} \frac{q_i q_j}{ r_{ij}}, & r_{ij} \leq r_{\text{coul}}~,\\
      0~, & r_{ij}>r_{\text{coul}}~,
      \label{Coul}
\end{cases}
\end{equation}
where $r_{\text{coul}}$ is the cutoff value for the long-range Coulomb interaction.  $\varepsilon$ and $\varepsilon_0$ are the dielectric constant of the solvent and the vacuum permittivity, respectively. Eq.~\eqref{Coul} is evaluated in LAMMPS\cite{LAMMPS} using the particle-particle-particle-mesh (PPPM) method, where short-range interactions are calculated directly in real space and long-range contributions are calculated in reciprocal space using a mesh-based solver. This is similar to evaluating the long-range part in the Fourier space using an Ewald summation technique. A real-space cutoff of 50$\sigma$ and a relative force accuracy of $10^{-5}$ are employed.  A Langevin thermostat is used to maintain the reduced temperature at $T^* = k_B T / \epsilon = 1$  between MC moves. The length scale and the energy are measured in units of $\sigma$ and  $\epsilon$, respectively. Values of mass $m$, energy $\epsilon$ and $k_B$ are taken as  $1$. With these, the time-step of integration $\delta t =0.005$   is taken in units of $\tau =\sqrt{m\sigma^2/\epsilon}$. The radius of the NP is kept fixed at $R=5\ell_B$  throughout the work and its position at $\vec{r}_{\text{NP}=}(100.0,100.0,33.33)$.

The electrostatic screening in the solution is characterized by the Debye length \cite{Hckel},
\begin{equation}
\lambda_D = \sqrt{\frac{\varepsilon\varepsilon_0 k_B T}{n e^2}},
\end{equation}
where $n$ is the number density of monovalent charge carriers. In our simulations, $n$ is related to the concentrations of positive and negative ions, i.e., $pI_p$ and $pI_m$, respectively, by $n = N_A\times(10^{-pI_p} + 10^{-pI_m})$ (with concentrations in mol/L converted to m$^{-3}$). Since $\lambda_D \propto n^{-1/2}$, a high value of $pI$ (low ion concentration) provides long-range PE-NP attraction, while a low $pI$ (high ion concentration) strongly screens it. Throughout the paper, we refer to the low and high concentration cases by their respective $pIp$ amd $pIm$ values.

\begin{table}[t!]
\centering
\caption{LJ interaction parameter $\Delta$ between different particle types, categorized as small (mobile ions, surface sites, and monomers) with radius $r=\sigma/2$ and NP with radius $R$.}
\label{tab:table1}
{\setlength{\tabcolsep}{4pt}\setlength{\extrarowheight}{3pt}
\begin{tabular}{|c|c|}\hline
\parbox[c]{3.8cm}{\centering\textbf{Interaction between}\\
\textbf{Particles}\vspace{3pt}} &
\parbox[c]{1.2cm}{\centering\textbf{Value of $\Delta$}\vspace{3pt}}\\ \hline
\parbox[c]{3.8cm}{\centering NP with small ions and monomers ($R\leftrightarrow r$)} &
$R-r$\\ \hline
\parbox[c]{3.8cm}{\centering Ions or monomers with other ions and monomers ($r\leftrightarrow r$)} &
$0$\\ \hline
\end{tabular}}
\end{table}

\section{Results}
\label{sec:result}
\begin{figure*}[t]
\includegraphics[width=0.8\textwidth]{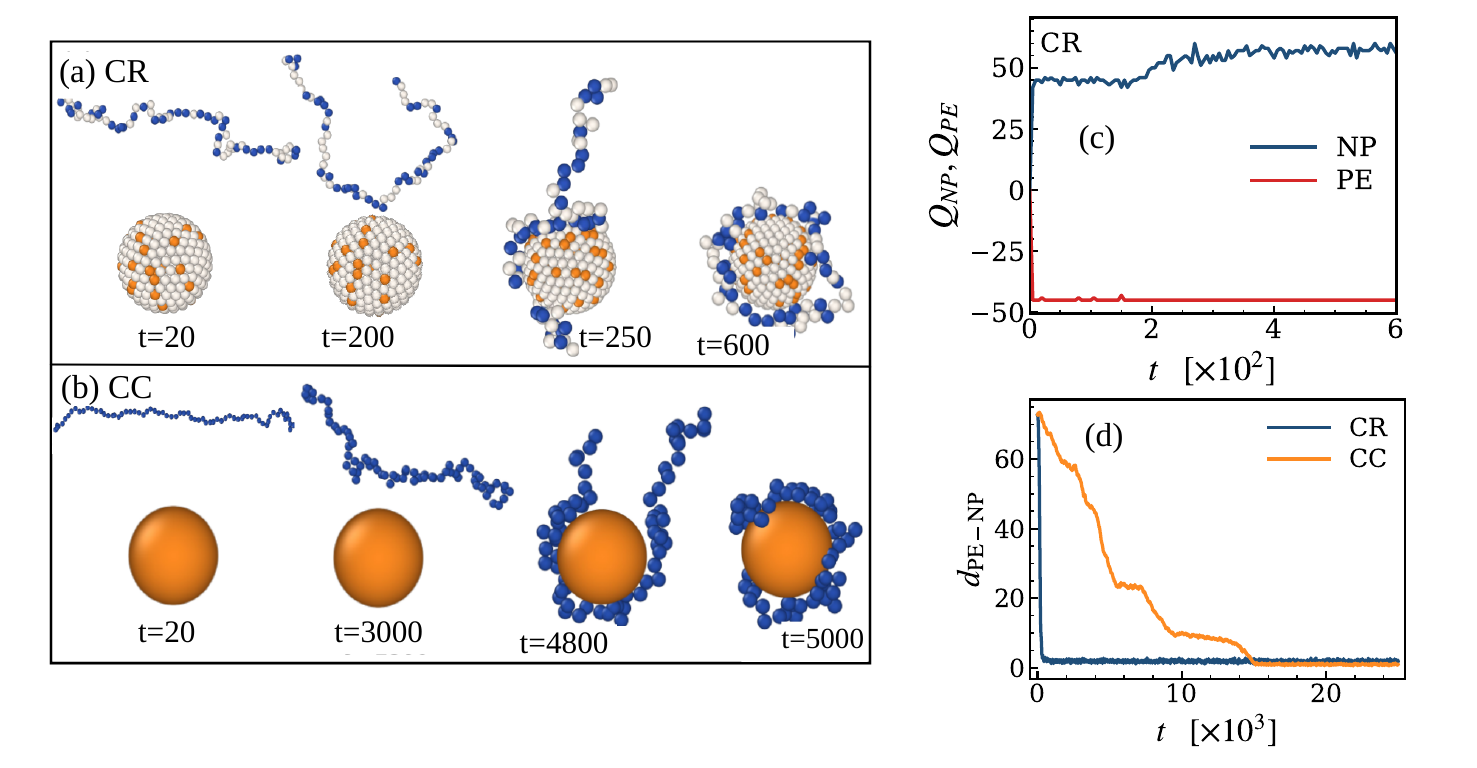}
\caption{Time-evolution snapshots comparing the (a) charge regulation (CR) and (b) constant charge (CC) models at low ion concentration, $pI_p = pI_m = 5$. The time of each snapshot is given below it. The snapshots show the adsorption of a PE of length $N = 70$ onto a nanoparticle (NP) of radius $R = 5\ell_B$. (c) Time evolution of the charges $Q_{\text{PE}}$ and $Q_{\text{NP}}$ for the CR case. (d) Distance $d_{\text{PE-NP}}$ versus $t$ for the CR and CC cases; the decrease of $d_{\text{PE-NP}}$ to its minimum value corresponds to the adsorption of the PE onto the NP. In the CR case, the NP and the PE carry ionizable basic and acidic groups, respectively, with $pK_b = 3$ and $pK_a = 3$. Solvent ions are not shown in the snapshots for visual clarity.}
\label{Snap_low}
\end{figure*}
\subsection{Low-concentration case}
We begin with the case of low salt concentration, with $pIp = pIm = 5$. First, we show the time evolution snapshots of a PE of length $N = 70$ during its interaction with the NP for both CR and CC case in Figure~\ref{Snap_low}(a)-(b). The corresponding system parameters are $pK_a = pK_b = 3$. In both CC and CR cases, starting with an extended PE at $t=0$, we present snapshots at various times until the adsorption of PE onto the surface of the NP becomes quite evident.  The time evolution snapshots in the CR show that for both PE and NP their charged sites are continuously relocating, helping to strengthen the kinetics of adsorption process, i.e., attraction between PE–NP and facilitating PE wrapping around the NP surface. Such dynamic charge redistribution results in faster conformational relaxation and  adsorption compared to those in the CC case. The acidic monomers of the PE chain repel each other and force the monomers to spread uniformly across the surface of the NP.


To elucidate the dynamics of PE adsorption, the PE--NP interaction can be monitored by tracking the time-dependent charge fluctuations in the CR case as well as the separation between them. In Fig.~\ref{Snap_low}(c) we show the variation of $Q_{\text{NP}}$ and $Q_{\text{PE}}$ versus $t$ for the CR case. At early times, the magnitude of the net charge on both the PE and the NP increases rapidly, as their surface sites (de)ionize in response to the local environment. As the PE approaches and adsorbs onto the NP surface, the ionization state of the NP sites fluctuates strongly, driven by the electrostatic interaction with the PE. The resulting gain in electrostatic interaction energy can compensate for the chemical free-energy cost associated with shifting the acid--base equilibria toward a higher degree of ionization. As a result, the total free energy decreases and both components increase their charge in the adsorption state. Once the PE is adsorbed or very close to the NP surface, these fluctuations subside and $Q_{\text{NP}}$ and $Q_{\text{PE}}$ saturate to their steady-state values. However, $Q$ reaches a steady-value earlier than $d_{\text{PE-Np}}$ reaches its minimum. 
\begin{figure*}[t]
    \includegraphics[width=0.95\textwidth]{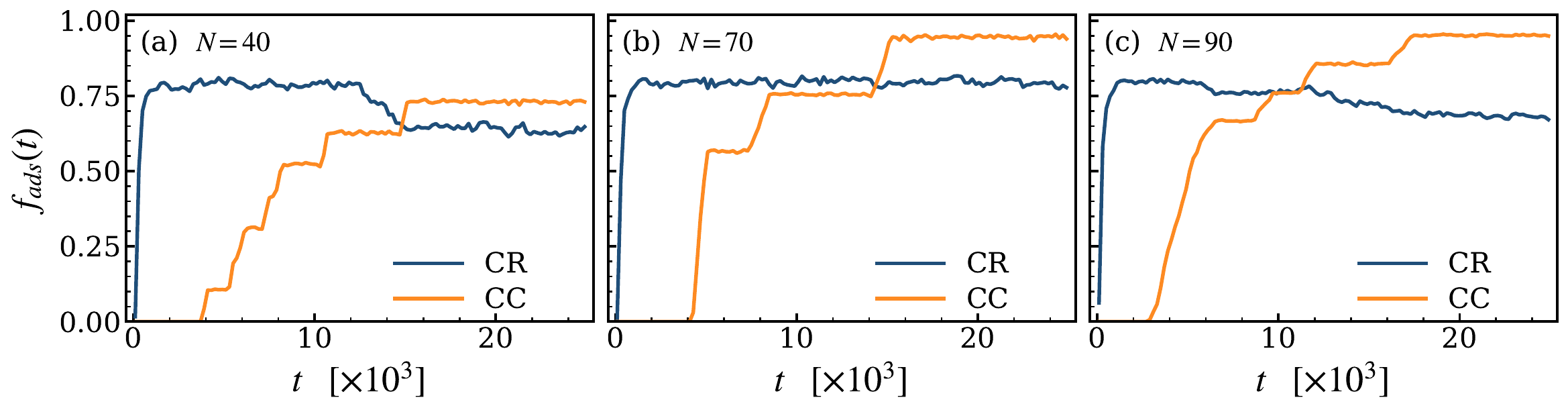}
    \caption{ Plots of adsorption fraction $f_{\text{ads}}$ for three different values of PE length for both CR and CC cases. (a)-(c) correspond to the data for different values of the PE length $N =40,70,90$. All data correspond to a fixed NP radius $R=5\ell_B$ with fixed value of the concentration $pIp = pIm = 5$. All data are averaged over $10$ independent simulations. In case of CC, step-wise increase of $f_{\text{ads}}$ is observed for all the chain lengths.}
    \label{lfad}
\end{figure*}

A quantitative analysis of the charge regulation effect is carried out by calculating the number of adsorbed monomers $N_{\text{ads}}$ of PE on the surface of NP. The number $N_{\text{ads}}$ is calculated whose positions are within a distance $3\sigma$ from the NP surface. The fraction $f_{\text{ads}}$  is calculated as $f_{\text{ads}}= N_{\text{ads}}/N$. Figure~\ref{lfad}(a-c) illustrates the time evolution of $f_{\text{ads}}$ for both CR and CC cases, for three different lengths of the PE, i.e., $N = 40, 70$, and $90$ with fixed values of $pK_a = pK_b = 3$. For any length $N$ of the PE, the value of $f_{\text{ads}}$ saturates asymptotically at a non-zero value for both CR and CC cases. Interestingly,  for the CR case, $f_{\text{ads}}$ increases very rapidly and saturates to a value close to $\approx 0.75$. On the other hand, for the CC case, $f_{\text{ads}}$ increases in a step-wise manner following some intermediate plateaus, reflecting slow dynamics of the PE during its gradual  and steady move towards the NP. For longer chain lengths, $f_{\text{ads}}$ appears to saturate to a higher value for CC than for the CR case, as seen from Figure~\ref{lfad}. Even though the dynamics is slower, the CC case provides a more stable complexation during the adsorption process. At low salt concentrations, the strong electrostatic attraction between the PE and NP compensates for the entropic penalty arising from the loss of translational and conformational degrees of freedom of the PE chain upon adsorption. This behavior is in agreement with the recent case of PE adsorption on a planar surface \cite{Yuan2024} and PE adsorption on two NPs \cite{Yadav2026}.

Our results show that the presence of CR significantly accelerates the interaction between PE and NP. This is attributed to the dynamic adaptability of the surface charge, which facilitates rapid electrostatic adjustment and accelerates the binding of PE on NP. However, contrary to previous studies on PE adsorption on planar surfaces\cite{Yuan2024} and cases involving two NPs \cite{Yadav2026}, here, the value of $f_{\text{ads}}$ in CC case remains somewhat higher than that in the CR approximation. This may be due to the fact that, in the CR case, only $50\%$  monomers of the PE and a very few charged sites ($50 \%$ of the overall surface) of NP have been allowed to undergo ion exchange with solution and the corresponding charge fluctuations.




\begin{figure*}
\includegraphics[width=0.95\textwidth]{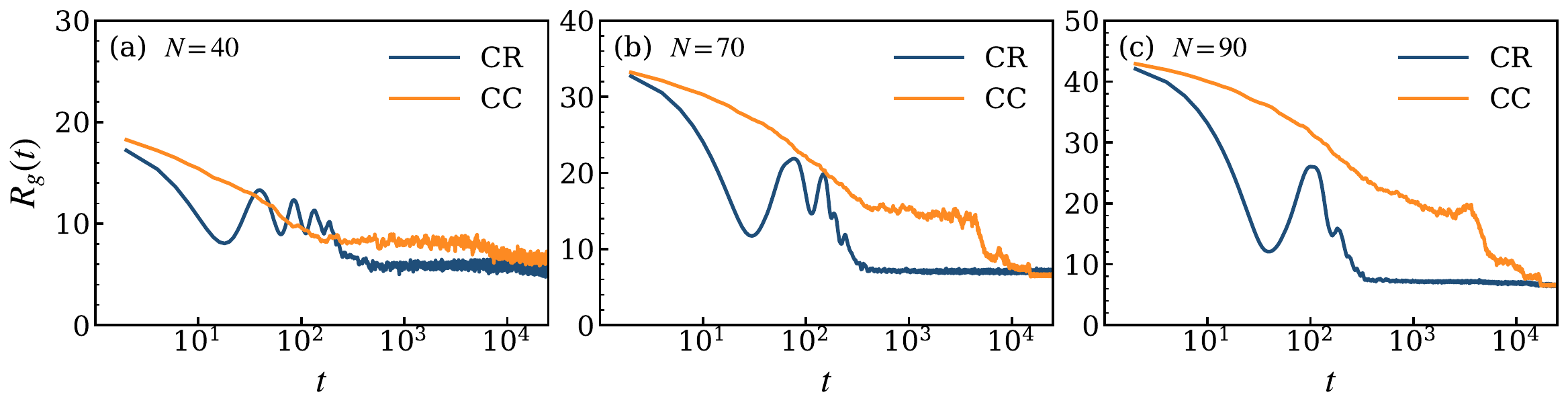}
\caption{ Plots of radius of gyration $R_g(t)$ versus $t$ for three different PE length $N =40,70,90$ and a fixed NP of radius $R=5\ell_B$ for both CR and CC cases. The concentration is fixed at $pIp = pIm = 5$. All data are averaged over $10$ independent runs. }
\label{lrg}
\end{figure*}
\begin{figure*}
\includegraphics[width=0.7\textwidth]{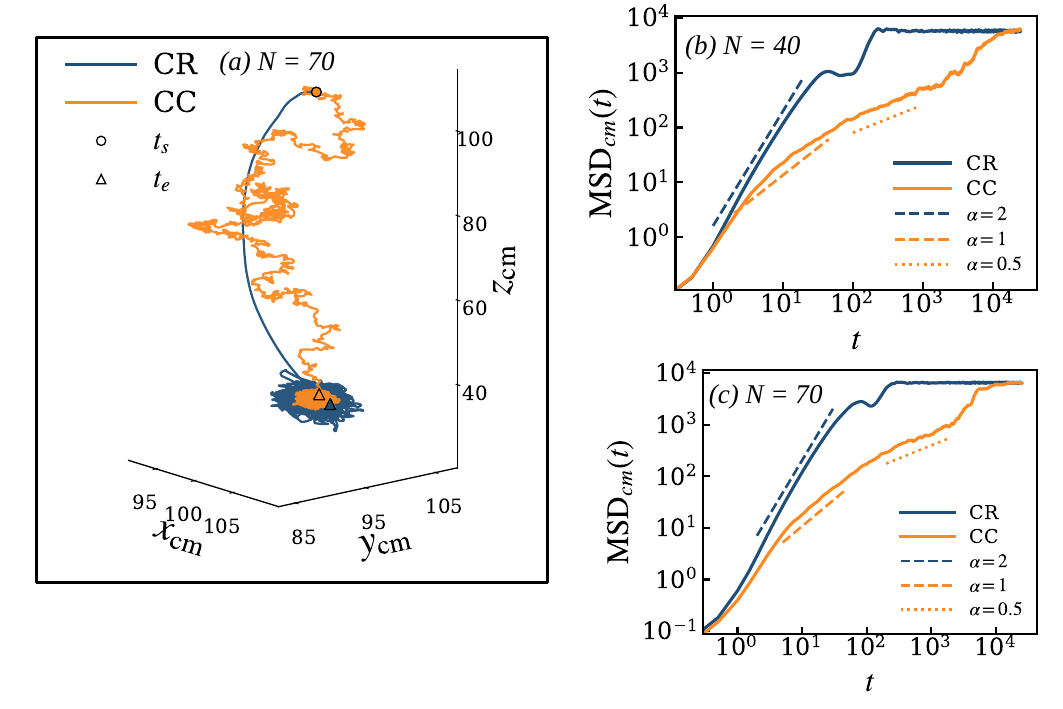}
\caption{(a) Center-of-mass trajectories of the PE for the CR and CC cases for $N = 70$; $t_s$ and $t_e$ mark the starting and final times of the trajectories. (b), (c) Mean-squared displacement of the center of mass of the PE, $\mathrm{MSD_{cm}}(t)$, versus $t$ for the CR and CC cases for two chain lengths, $N = 40$ and $70$, during interaction with an NP of radius $R = 5\ell_B$. The ion concentrations are fixed at $pI_p = pI_m = 5$ and the dissociation constants at $pK_a = pK_b = 3$. The dashed lines show power-law scaling $\mathrm{MSD_{cm}} \sim  t^{\alpha}$ with the indicated exponents, where $\alpha = 2$ corresponds to ballistic-like, $\alpha = 1$ to diffusive, and $\alpha < 1$ to sub-diffusive motion. The saturation of the MSD at late times indicates the adsorption of the PE onto the NP. MSD data are averaged over 10 independent simulation runs.}
\label{lmsd}
\end{figure*}

It is evident from the adsorption results in Fig.~\ref{lfad} that, in the CR simulations, the adsorption behavior do not depend much on the PE chain length. This is somewhat consistent with the fact that the adsorption of PE on oppositely charged surfaces exerts a neutralization effect, which appears to only slightly influenced by the length of the PE chains \cite{Wang2010}. In all these cases, PE prefers to adsorb on NP almost immediately. Overcharging or charge reversal can occur in the presence of very long chains or multiple PEs \cite{Vagharchakian2004, Dobrynin2001,Grosberg2002,Lenz2008}. In the CC case, the relative effect of the PE chain length becomes significant. The time required for PE adsorption onto the NP changes with the length of the PE chain. For $N=70$, the time required for adsorption is $\approx 1.5$ times that of $N=40$. This illustrates the complex interplay of electrostatic interactions; as the PE length increases, the repulsion among the monomers also increases. Once the PE approaches the NP, the attractive force between the PE and the NP becomes dominant, making the adsorption to happen.

In order to demonstrate the conformational changes of the PE we calculate its radius of gyration $R_g$ defined as
\begin{equation}
R_g = \Big\langle \bigg[\frac{1}{2N^2} \sum_{i,j} (\vec{r}_i - \vec{r}_j)^2\bigg]^{1/2} \Big\rangle ,
\end{equation}
where $\vec{r}_i$ stands for the position vector of the $i$-th monomer, and  $\langle \cdots \rangle$  denotes the average over independent initial conditions. $N$ corresponds to the number of monomers or length of the chain.  The time evolution of $R_g$ during the complexation process is shown in Fig.~\ref{lrg} for the same parameters as in previous figures. Data are shown for $N = 40, 70$, and $90$ while the radius of the NP is fixed at $R=5\ell_B$. With time, $R_g$ decreases from its starting maximum value for a rod-like conformation.  Subsequently, the PE moves towards the NP surface and saturates to a value of $R_g$ corresponding to its new equilibrium conformation after the complexation. In comparison to a free polymer, the presence of solvent ions and solvent-mediated interactions, along with competition between electrostatic monomer-monomer repulsion and Coulombic attraction between the PE and NP, can drive the PE into a variety of conformational states~\cite{Curk2022, Beyer2025, Radhakrishnan2025}. For both CC and CR, even though it finally wraps around the NP,  the time evolution of the $R_g$ are quite different.  In the CR scenario, due to dynamic and rapid adsorption process decay of $R_g$ is much faster compared to a steady decay in the CC case, leading to faster decrease of $R_g$. However, the fluctuations in conformations at intermediate times decrease with increasing chain length. This faster decrease in $R_g$ can be understood from the snapshots presented in Fig.~\ref{Snap_low}. 

The impact of chain length is more noticeable in the CC simulations.  $R_g$ relaxes slower for the larger chains as a result of the longer time taken for adsorption. Before adsorption, PE chains remain extended for longer times due to stronger monomer–monomer repulsion. In all CC cases, we observe that the longer PE chain  lowers the adsorption tendency. This is likely to be attributable to increased loss of entropy, which discourages PE adsorption in the early time regime\cite{Africo2024}. The time required for collapse increases significantly with the length of the PE chain, especially in the CC scenario, although both the CR and the CC systems eventually reach almost comparable equilibrium values $R_g$. 


This interesting conformational behavior between the CC and CR cases led us to probe the dynamics of the center of mass (cm) of the PE  during its interaction with the NP. From their respective trajectories, we calculate the corresponding mean square displacements (MSD) as,
\begin{equation}
\mathrm{MSD}_{\text{cm}}(t) = \left\langle \left[ \vec{r}_{\mathrm{cm}}(t) - \vec{r}_{\mathrm{cm}}(0) \right]^2 \right\rangle ,
\end{equation}
where $\vec{r}_{cm}(t)$ defines  the position of the cm of the PE.  In general MSD follows a power law behavior with time as 
\begin{equation}
    \text{MSD} \sim t^{\alpha}\,,
\end{equation}
where  the exponent $\alpha$ depends on the dynamics.  $\alpha=1$ and $2$ correspond to the diffusive and ballistic motions, respectively. $1 < \alpha <2$ corresponds to a super-diffusive motion, and $0<\alpha<1$ the particle exhibits a sub-diffusive behavior. MSD plots provide important information on the dynamics and on the possibility of interaction.


First we discuss the CC case. After a very early transient ballistic regime, at intermediate times $\text{MSD}_{\text{cm}}(t) \sim t$ and $\text{MSD}_{\text{cm}}(t) \sim t^{0.5}$, corresponding to diffusive and sub-diffusive motion of the PE. This is also somewhat clear from the  trajectory seen in Fig.~\ref{lmsd}(a) where $t_s$ and $t_e$ defines the starting and the end times, respectively. The sub-diffusive nature is linked with the weak electrostatic interaction between the PE and the NP. Even though the PE and NP are charged oppositely, the interaction becomes weak due to Debye screening or  thermal fluctuation of the monomers.  During the sub-diffusive motion, the conformations of the polymer change slowly,  as observed from  the decay of $R_g(t)$ shown in Fig.~\ref{lrg}. For all the chain lengths the exponent $\alpha$ remains  $\approx 0.5$  depicting similar sub-diffusive dynamics for all the chain lengths. However, after an intermediate sub-diffusive growth, when the distance between PE and NP becomes small (as observed from Fig.~\ref{Snap_low}(d)), the dynamics get slightly faster before it saturates which corresponds to the adsorption. The chain length also seems to have a minimal effect on the overall dynamics.
\begin{figure*}
\includegraphics[width=0.8\textwidth]{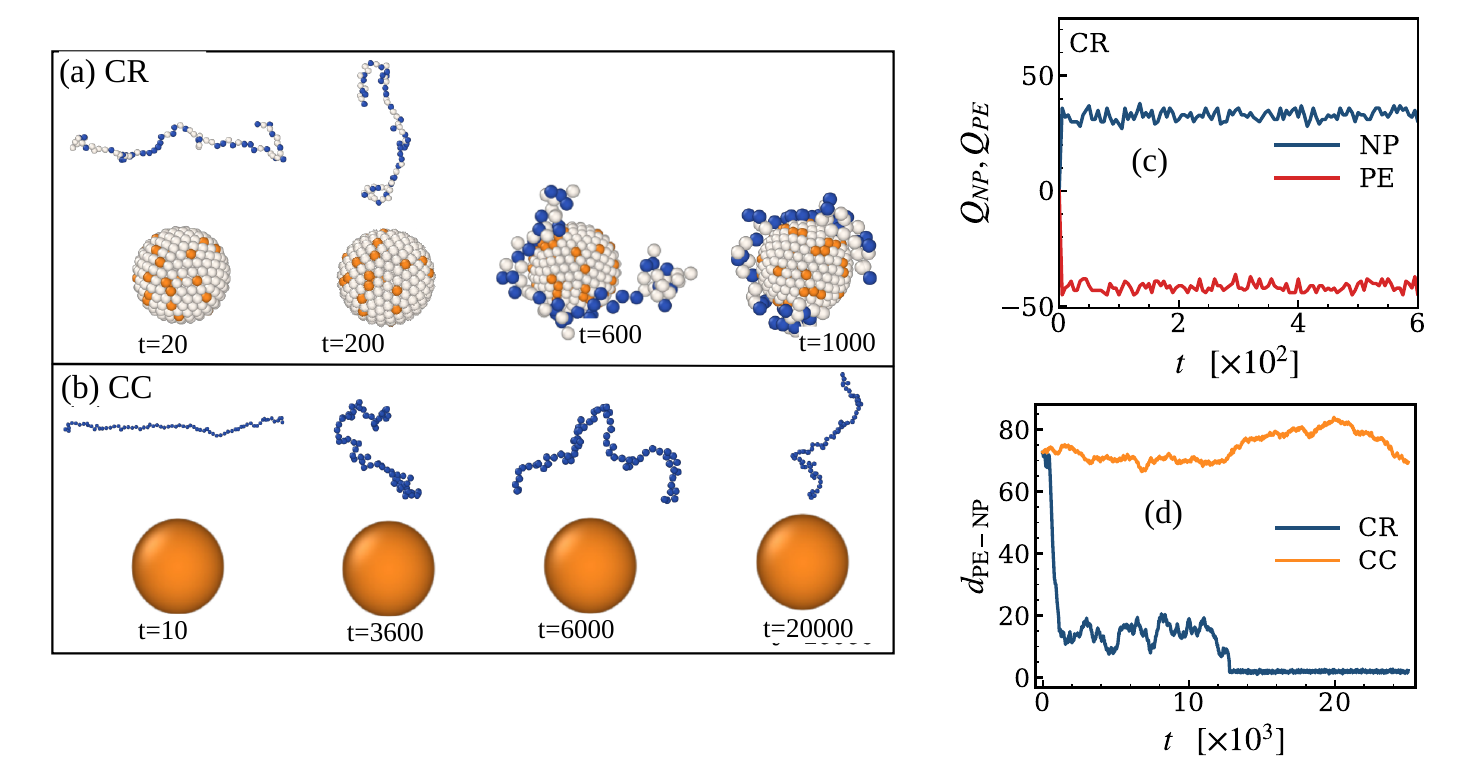}
\caption{Time-evolution snapshots comparing the (a) charge regulation (CR) and (b) constant charge (CC) models at high ion concentration, with $pI_p = pI_m = 3$. The corresponding time for each snapshot is given below it. The snapshots show the adsorption of a PE of length $N = 70$ onto a nanoparticle (NP) of radius $R = 5\ell_B$. (c) Time evolution of the charges $Q_{\text{PE}}$ and $Q_{\text{NP}}$ for the CR case. (d) Distance $d_{\text{PE-NP}}$ versus $t$ for the CR and CC cases. In the CR case, the NP and the PE carry ionizable basic and acidic groups, respectively, with $pK_b = 3$ and $pK_a = 3$. Solvent ions are not shown in the snapshots for visual clarity.}
\label{Snap_high}
\end{figure*}

On the other hand, for the the CR case, $\text{MSD}_{\text{cm}}(t) \sim t^2$ indicates the ballistic nature of its dynamics. Such ballistic behavior of the PE can be possible if all monomers move in similar direction or if there is a strong attractive force acting on the polymer. Unlike the case of active polymer where velocity alignment is the primary interaction, here,  a strong electrostatic attractive force between the PE and the NP drives ballistic-like dynamics. Both of them, being charge regulated, can adjust their surface charges to maintain strong interaction between them. As seen from the trajectory, in the CR case, the PE rapidly anchors onto the NP and tends to wrap around it by maximizing the number of PE–NP contacts, driven by the favorable thermodynamic attraction to minimize the corresponding free energy. Such wrapping up occurs when the interaction energy gained per monomer upon adsorption is sufficiently large to overcome the entropic penalty associated with formation of  a compact structure around the NP \cite{Africo2024}. After the fast ballistic approach, $\text{MSD}_{\text{cm}}(t)$ exhibits a short plateau followed by a sudden increase  before it saturates. However, with increasing PE length, the MSD data become smoother, and the sudden jump becoming less pronounced. This can be due to the fact that  with increasing PE length the interaction with the NP becomes more isotropic, which governs a much smoother dynamics of the PE towards its adsorption around the NP. However, after adsorption occurs, the monomers minimally fluctuate around their mean positions, and thus the MSD saturates. For our considered chain lengths, the time-scale at which the MSD reaches its saturation which also corresponds to the $R_g$ settling to its steady-state value are more or less similar. We also verified this for N = 90 and observed similar behavior for both CC and CR.
 


\begin{figure*}
    \includegraphics[width=0.95\textwidth]{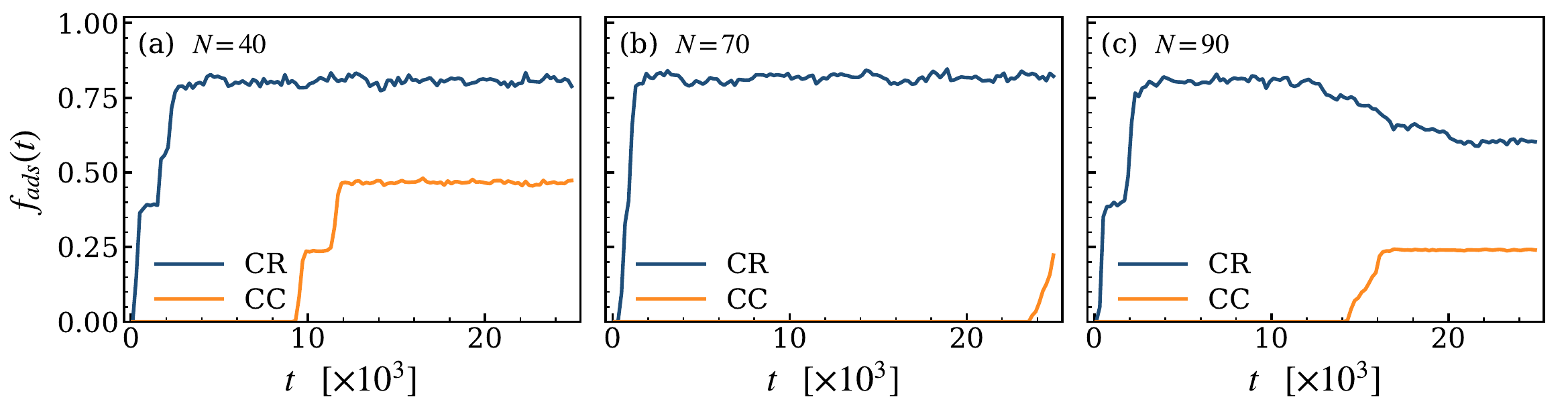}
    \caption{ Plots  showing the adsorption fraction $f_{\text{ads}}$ versus $t$ for both the CC and CR cases during interaction between the NP and PE. (a)-(c) correspond to three different values of PE length $N =40,70,90$ interacting with a fixed radius of the NP $R=5\ell_B$. The concentration $pIp = pIm = 3$ are fixed and higher in comparison to the data shown in Fig~\ref{lfad}. The NP-PE surface is characterized by surface dissociation constants $pK_a = pK_b = 3$.  Data are averaged over $5$ independent simulations. For the CC case, the representative snapshots for each value of $N$ close to the corresponding $f_{\text{ads}}$ data. } 
    \label{hfad}
\end{figure*}


\begin{figure*}[ht]
\includegraphics[width=0.95\textwidth]{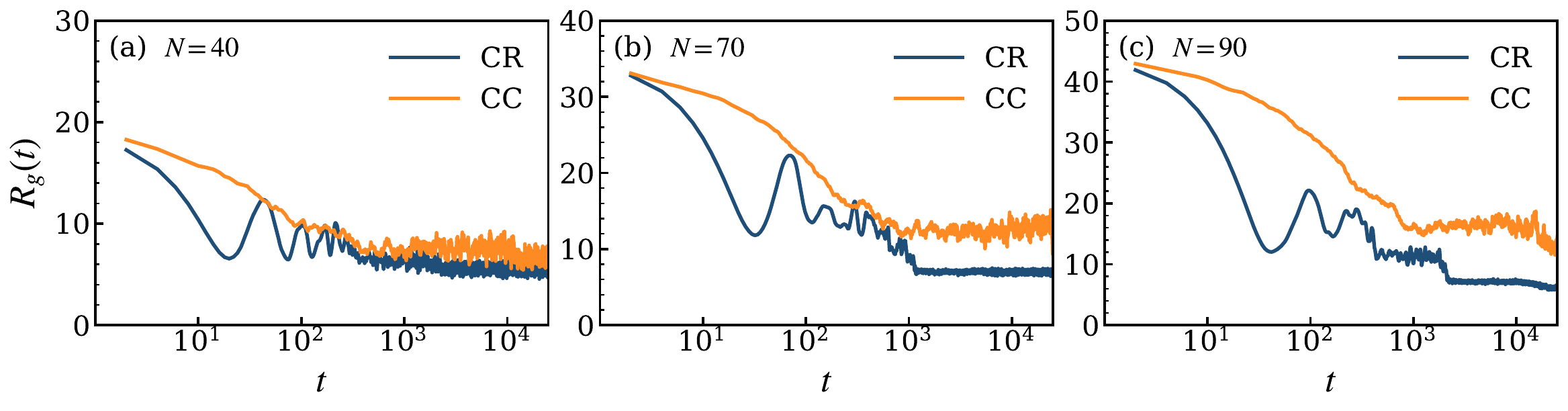}
\caption{ Plots show time evolution of the radius of gyration $R_g(t)$ for the PE chain for both CC and CR cases. (a)-(c) correspond to the data for three different values of PE length $N =40,70,90$ interacting with  a fixed NP radius $R=5.0\ell_B$. The concentration is fixed at $pIp = pIm = 3$. The NP-PE surface is characterized by surface dissociation constants with values $pK_a = pK_b = 3$.  All data are averaged over $5$ independent simulations. Data of $R_g$ for the CR case for $N=70$ and $90$ saturates at a higher value compared to CC case.} 
\label{mrg}
\end{figure*}

\begin{figure*}
\includegraphics[width=0.7\textwidth]{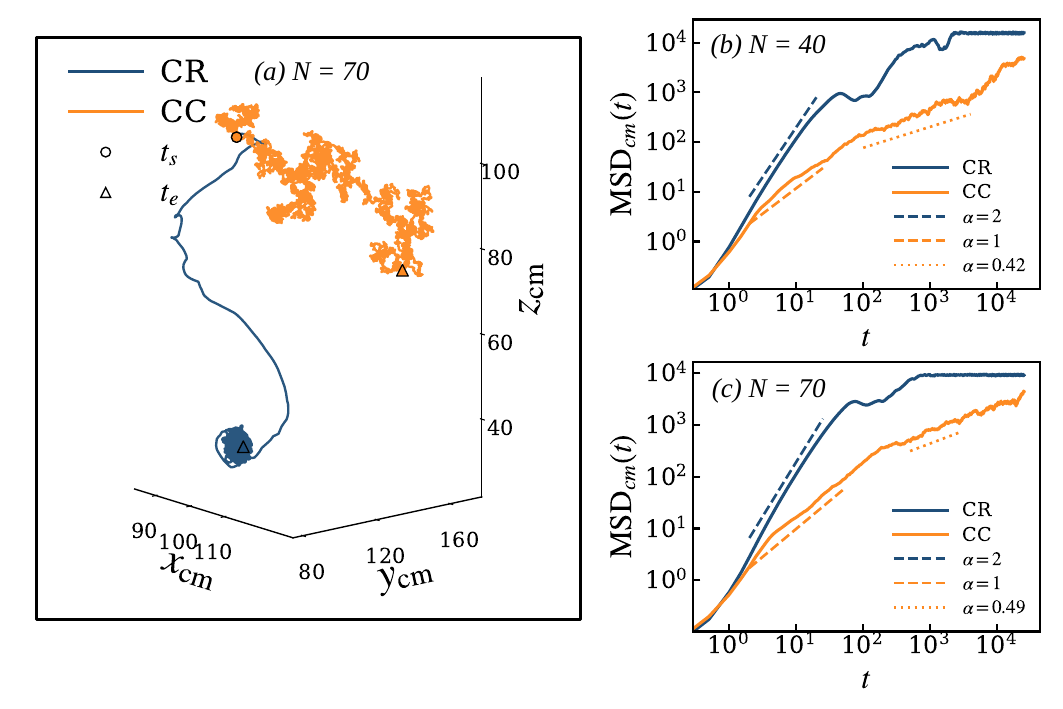}
\caption{ (a) Center-of-mass trajectories of the PE for the CR and CC cases for $N = 70$; $t_s$ and $t_e$ mark the starting and final times of the trajectories. (b), (c) Mean-squared displacement of the center of mass of the PE, $\mathrm{MSD_{cm}}(t)$, versus $t$ for the CR and CC cases for two chain lengths, $N = 40$ and $70$, during interaction with an NP of radius $R = 5\ell_B$. The ion concentrations are fixed at $pI_p = pI_m = 3$ and the dissociation constants at $pK_a = pK_b = 3$. For the CR case, the saturation of the MSD at late times indicates the adsorption of the PE onto the NP.  Mentioned values of $\alpha$ correspond to scaling behavior shown by the different dashed lines. MSD data are averaged over 5 independent simulation runs.}
\label{hmsd}
\end{figure*}


\subsection{High Concentration case}
Now we move to the higher salt concentration case with  $pIp = pIm = 3$. The value of $pH$ and other system parameters are identical to those used earlier. Similar as before, Fig.~\ref{Snap_high}(a)-(b) represents the time-evolution snapshots for a PE chain of length $N = 70$ interacting with an NP of size $R=5\ell_B$ under both CR and CC conditions. Compared to the low-concentration case, the increased ionic concentration modifies the electrostatic environment and enhances the screening effects \cite{Avni2018,AVNI201970,tomar17,Kandari2026}, thereby influencing the conformational changes and adsorption dynamics of the PE chain. 

For the CC case, within our simulation time frame, there is no discernible adsorption of the PE  onto the NP surface, as seen from the final time snapshot  at $t=20000$. For most of the runs, the PE remains largely disengaged from the NP throughout the simulation time. This behavior can be attributed to enhanced electrostatic screening at higher concentrations, which weakens the attractive interaction between the PE and the fixed surface charges of NP, consistent with previous studies \cite{Forsman2012, Shafir2003, Yadav2026}. In this case, charge regulation plays an important role. For the CR case, even though only a small fraction of sites of the NP and the number of monomers of the PE take part in charge regulation, this helps the PE to finally get adsorbed on the NP. As seen from the corresponding times, the complexation process  takes longer in comparison to the lower concentration case. Fig.~\ref{Snap_high}(c) shows the charges $Q_{\text{PE}}$ and $Q_{\text{NP}}$ developed over time for the PE and NP, respectively. Like in the previous cases, the saturation is reached quite rapidly. In Fig.~\ref{Snap_high}(d) the distance $d_{\text{PE}-{\text{NP}}}$ is shown for both CR and CC cases. For the CC case, $d_{\text{PE}-{\text{NP}}}$ remains at a higher value $\approx 80$ in units of $\sigma$ indicating almost no adsorption of the PE.\\



Figure~\ref{hfad} shows the corresponding time evolution of the adsorbed monomer fraction $f_{\text{ads}}$ versus $t$ for various PE chain lengths. In CR simulations, regardless of the chain length, $f_{\text{ads}}$ changes rapidly over time and quickly reaches a saturation value somewhat similar to the low concentration case, as reported in previous studies as well \cite{Wang2010}. However, the time taken to reach saturation is longer than in the low concentration case. Due to CR, i.e., the dynamical modification of the charge distribution of the PE and NP, the interactions persist even under highly screened environments. In contrast, in the CC simulations, no discernible adsorption is observed within the simulated time scale. Nevertheless, a very small nonzero $f_{\text{ads}}$ appears at late times for $N=40$, $70$, and $90$. This stems from adsorption events taking place only for a few independent runs.  Thus, while averaged over all the runs, $f_{\text{ads}}$ yields a small nonzero value at late times.  Given the limited number of runs, we do not make any conclusive remark regarding  the chain length dependence on $f_{\text{ads}}$ in the CC case. Should such a dependence exist, it could arise from a competition between the stronger PE--NP attraction provided by additional monomers and the increased number of unpaired monomers, which intensifies the repulsion between like-charged monomers. These results indicate that the disparity between CR and CC adsorption dynamics is more noticeable at high salt concentration, underscoring the crucial role of CR in promoting PE adsorption under strongly screened conditions.


The charging behaviour of NP and PE is shown in  Figure~\ref{Snap_high} (c). For 1 mM salt, the charges on the NP and PE increase significantly due to the screening effect, and both species regulate their charges. The PE itself becomes nearly ($50 \% $) charged in all three PE chain lengths cases. Upon adsorption, a cooperative CR mechanism strongly enhances ionization of both the PE and the NP. Under CR conditions the system lowers its free energy through cooperative ionization between PE and a single NP. This adaptive charging favors localized adsorption. Once PE adsorbs onto the NP its charges saturates at the maximum allowed by its degree of polymerization. 


Now we focus on the conformational properties and the dynamics of the PE as it approaches the NP.  In Fig.~\ref{mrg}(a)-(c) we show the time evolution of $R_g$ of the PE for both CR and CC cases, for three different values of  $N$. Different stages of decay of $R_g$ observed in the CR scenario are somewhat similar to the low concentration case in Fig.~\ref{lrg}. At intermediate times, the PE shows larger variation in its conformations as $d_{\text{PE}-\text{NP}}$ also shows considerable fluctuations.  However, in comparison to the lower ion concentration case with  $pIp = pIm = 5$, the time required for its adsorption is higher for all lengths of PE. On the other hand, for the CC case, the impact of solvent ions is more noticeable. As usually the adsorption does not occur, $R_g$ remains at a higher value (see comparison with Fig.~\ref{lrg}). This behavior indicates that under strongly screened conditions how the charge-regulation dynamically modify its charge state to retain enough attractive interaction for the PE adsorption.

In Fig.~\ref{hmsd}(a), we plot typical trajectories of the center-of-mass of the PE for both CR and CC cases, for $N=70$. Starting from same point of $\vec{r}_{\text{cm}}$ of the PE at $t=0$, the final positions are different. For CR, as adsorption happens, following a more directed motion, the trajectory ends up on the surface of the NP. For the CC case, the trajectory follows a diffusive-like motion, and does not end up on the surface of the NP. Corresponding plots of $\text{MSD}_{\text{cm}}$ versus time are shown in Fig.~\ref{hmsd}(b)-(c) for different values of PE chain length, $N=40$ and $70$.  At early times, with increasing salt concentration, CR simulations reveal no qualitative change in the MSD scaling with increasing salt concentration. $\text{MSD}_{\text{cm}}$ shows $ \sim t^2$ scaling for quite longer period. However, as the PE approaches NP, the dynamics becomes slower with MSD deviating from a ballistic scaling before finally showing its saturation. This intermediate slower dynamics was absent in the lower concentration case.  On the other hand, $\text{MSD}_{\text{cm}}$ for the CC case shows diffusive followed by a sub-diffusive with $\alpha$ slightly lower than   $0.5$ for a very longer period until our simulation time. And, due to no adsorption, the MSD does not approach a constant value. It is important to note that in this high salt concentration the CR condition has a more significant impact on the PE dynamics for the formation of complexation than the  CC condition.

\section{Summary and Future Outlook}
\label{sec:con}
Effects of charge fluctuations are quite fundamental and pervasive in many biological and technologically relevant systems. The importance of charge fluctuation is particularly pronounced in systems where the solution conditions (PH, salt density, valencies) and presence of other charged moieties can be precisely tuned to control the ionization sate of charge decorated particles. 
In this work, we examine the ionic strength driven surface charge modification of a nanoparticle (NP)  during its interaction with an oppositely charged polyelectrolyte (PE). The effect of charge regulation (CR) is studied using a hybrid Monte Carlo/molecular dynamics simulation technique. Results related to the conformation and adsorption of PE for the CR case are compared with those of a similar system consisting of NP and a PE having their charge as constant. Our results indicate that at low salt concentrations, for both CC and CR the adsorption of the PE occurs on the NP. Even a small number of CR sites on the NP surface accelerates this process, and PE follows fundamentally different kinetic pathways compared to CC paradigm. In comparison to a faster adsorption for CR, in the CC case the adsorption happens in a step-wise manner. Interestingly, for longer chains the adsorption looks better for the CC case, i.e., the complexation remains more stable. Furthermore, the conformations  are looked at from the radius of gyration $R_g$. As expected, $R_g$ decreases much faster for CR, compared to a   more gradual decrease for the CC case. This fact is related to the dynamics of the PE as it starts interacting with the NP in presence of salt ions. Also, the effect of chain length on the conformations during wrapping on the NP is more prominent in case of CC simulations. In comparison to the ballistic-like dynamics of the PE due to strong Coulombic interactions in the CR case, the PE transport mechanism is diffusive followed by a late-time sub-diffusive behavior in the CC scenario. During adsorption, a cooperative CR mechanism rapidly enhances the ionization of both the PE and NP, resulting in faster adsorption dynamics.   
However, at a high salt concentration of 1 mM, our results show significant differences between the CC and the CR scenario. Whereas  in the CR case, the PE shows ballistic dynamics and rapid adsorption, in the CC case, one observes a prolonged  sub-diffusive transport with a much lower probability of complexation. Varying the PE length yields only minor quantitative changes, showing that the charge fluctuation mechanism and the ionic environment play the dominant role in determining adsorption behavior. A charge-regulated NP surface can thus maintain sufficient attraction to sustain PE adsorption and compaction by dynamically altering its charge state, even under strong screening. The pronounced differences observed between CR and CC therefore do not originate from differences in mean charge but rather from the presence of charge fluctuations and spatially heterogeneous charge distributions that are intrinsic to the CR approach and absent in CC.

Our results reveal that at high salt concentration, CR ensures the adsorption of the PE on the NP and leads to a final equilibrium state of complexation between them. It would be interesting to systematically vary the size of the NP and investigate the effect of salt concentration on the adsorption process. In this work, $50\%$ of the total sites of both PE and NP are considered to be involved in charge fluctuations. However, in realistic systems, the fraction of ionizable sites depends on the local environment, including pH and the local electrostatic potential \cite{Anhesini2026}.  As an ongoing work, we are varying the pH together with the fraction of fluctuating sites, which we expect to yield a rich adsorption behavior as an interplay between CR and electrostatic screening. In addition, while analyzing the charge capacitance of such an PE-NP system, we see a non-monotonic behavior as a function of these parameters. Finally, in a realistic scenario one can consider a semi-flexible polymer, for which the effects of bending rigidity on conformations, knotted structures, and free-energy minimization are also worth investigating. On a similar scenario, considering block copolymers with different blocks having different CR behavior can also be interesting to look at.

A straightforward extension can be to consider multiple PEs  interacting with NP which could mimic several experimentally relevant scenarios, such as layer-by-layer adsorption of differently charged PEs and competitive adsorption of PEs with different charge densities, chain lengths, flexibilities, and adsorption strengths. In such a scenario, it would be particularly interesting to investigate how inter-chain interactions modify the conformational and dynamical properties of individual PEs.  Finally, in a more biologically relevant scenario, one can consider the case of active polymer which in a charge neutral condition also shows a persistent directed  motion \cite{majumder_prmat,paul2021motion}. Thus, one can consider the comparative scenario between both CC and CR conditions with the PE and NP adsorbed state but also the adsorption kinetics and transport dynamics. These findings underscore the relevance of adding CR effects when modeling PE-NP interactions under realistc solution conditions.


\section{Acknowledgment}
RK acknowledges the fellowship provided by the Ministry of Education (MoE), Government of India. SK acknowledges the financial support received from the Anusandhan National Research Foundation, India ( EEQ/2023/000676) and IIT Jodhpur for a research initiation grant (I/RIG/SNT/20240068). SP acknowledges University of Delhi for providing financial assistance through the Faculty Research Programme under Grant-IOE (Ref. No. IOE/2024-25/12/FRP).


\FloatBarrier

\bibliography{Manuscript} 

\end{document}